\documentstyle[aps,floats,epsf,psfig,twocolumn]{revtex}

\begin{document}

\twocolumn[\hsize\textwidth\columnwidth\hsize\csname
@twocolumnfalse\endcsname

\title{Effective Kondo model for a trimer on a metallic surface}
\author{A. A. Aligia}
\address{Centro At\'omico Bariloche and Instituto Balseiro, 
Comisi\'on Nacional de Energ\'{\i}a At\'omica, 8400 Bariloche, Argentina}
\maketitle

\begin{abstract}
I consider a Hubbard-Anderson model which describes localized orbitals in
three different atoms hybridized both among themselves 
and with a continuum of
extended states (ES). Using a generalized Schrieffer-Wolf transformation, I derive an
effective Kondo model for the interaction between the doublet ground state
of the isolated trimer and the ES. For an isoceles trimer with distances $a$,$l$,$l$
between the atoms, the Kondo temperature is very small for $l<a$ and has
a maximum for finite $l>a$ when $a$ is small. 
The results agree with experiments for a Cr trimer on Au(111).

\end{abstract}


] 

The study of many-body phenomena in nanoscale systems has attracted much
attention in recent years. Progress in nanotechnology has made it possible to
construct nanodevices in which the Kondo physics is clearly displayed \cite
{gold,cron,wiel}. Scanning tunneling spectroscopy has made it possible to probe
the local density of states near a {\em single} impurity and Fano
antiresonances have been observed for several magnetic impurities on metal
surfaces \cite{mad,man,naga,knorr}. These antiresonances observed in
the differential conductance, reflect a dip in the  spectral
density of conduction states caused by the Kondo effect \cite{revi}. Furthermore,
corrals built on the (111) surface of noble metals or Cu have been used to
project the spectral features of the Fano-Kondo antiresonance (FKA) to
remote places \cite{man,revi}. The observed Fano line shapes for one
magnetic impurity on these surfaces have been reproduced by many-body
calculations \cite{revi,meri,lin}.

The situation is however different for Cr trimers on the Au(111) surface 
\cite{jan}. Experimentally, no FKA is observed for a linear trimer, or a
compact equilateral one in which the distance between any two Cr adatoms is
believed to be the same as the Au-Au nearest-neighbor one ($a=2.88$ \AA ).
For a single Cr adatom, no FKA is observed either. This implies that in 
these cases, if a Kondo effect originating from localized moments exists, its
characteristic temperature $T_{K}$ is less than the experimental temperature
(7 K). On the other hand, surprisingly, for an isosceles trimer with one
side of length $a$ and two sides of length $\sim 5.2$ \AA , the spectrum
does display a FKA with a half width $T_{K}=50\pm 10$ K. This spectrum is
similar to the one observed for one Co impurity on Cu(111) \cite{man,knorr}.

The origin of this puzzling dependence of $T_{K}$ of the Cr trimer with
geometry is still unclear, in spite of recent theoretical efforts \cite
{kuda,laza,note2}.

In this Letter, I start from a Hubbard-Anderson Hamiltonian that describes
three Cr adatoms interacting with a continuum of extended states (ES). Using
a generalized Schrieffer-Wolff transformation (SWT) \cite{sw}, the model is
mapped into the usual one-impurity Kondo model. The resulting effective
Kondo interaction averaged over the Fermi energy $J$ depends on the geometry
of the trimer. The ensuing Kondo temperature is consistent with the above
experimental observations \cite{jan}.

The form of the Hamiltonian is suggested by experimental evidence and
previous theoretical work. The line shape observed for the isosceles Cr
trimer with $T_{K}\sim 50$ K is rather symmetric, while calculations using
Wilson's renormalization group for the degenerate Anderson model obtain a
strongly asymmetric Kondo resonance \cite{zhu}. This suggests that there is
only one relevant 3d orbital. The surface potential breaks the degeneracy of
the d orbitals into two orbital doublets with angular momentum projection $%
m=\pm 1$ and $m=\pm 2$ and one orbital singlet with $m=0$ 
(the 3d$_{3z^{2}-r^{2}}$ orbital). Therefore the latter, singly occupied, is the
candidate for the formation of the Kondo state. A similar conclusion has
been reached for Co atoms on Cu(111) \cite{revi,meri,lin}. The absence of a
Kondo effect for Cr  \cite{jan} or Co dimers on Au(111) at interatomic
distances less than 6 \AA\ \cite{chen} suggests a strong interatomic hopping
leading to a spin singlet ground state. Due to the symmetry breaking of the
surface potential, the  3d$_{3z^{2}-r^{2}}$ orbital of the adatom 
hybridizes with
the 4s orbital, and then one expects a decay $r^{-1}$ with distance $r$ of the
hopping between hybrid localized orbitals \cite{harr}. 
Finally the fact that $T_{K}<7$ K for a single Cr
adatom points to a small hybridization between localized and ES, justifying
the use of 
the SWT.

The Hamiltonian can be written as 
\begin{eqnarray}
H &=&H_{\text{Cr}}+H_{\text{Au}}+H_{V}\text{,}  \nonumber \\
H_{\text{Cr}} &=&E_{d}\sum_{i\sigma }d_{i\sigma }^{\dagger }d_{i\sigma
}+Ud_{i\uparrow }^{\dagger }d_{i\uparrow }d_{i\downarrow }^{\dagger
}d_{i\downarrow }  \nonumber \\
&&-\sum_{\langle ij\rangle \sigma }t_{ij}(d_{i\sigma }^{\dagger }d_{j\sigma
}+\text{H.c.}),  \nonumber \\
H_{\text{Au}} &=&\sum_{{\bf k}\sigma }\epsilon _{{\bf k}}c_{{\bf k}\sigma
}^{\dagger }c_{{\bf k}\sigma },  \nonumber \\
H_{V} &=&\sum_{{\bf k}j\sigma }(V_{{\bf k}}e^{i{\bf k\cdot R}_{j}}d_{j\sigma
}^{\dagger }c_{{\bf k}\sigma }+\text{H.c.}).  \label{ham}
\end{eqnarray}
$H_{\text{Cr}}$ is a Hubbard model that describes the isolated Cr trimer,
with hopping that depends on the distance 
$r_{ij}=|{\bf R}_{i}-{\bf R}_{j}|$ between Cr adatoms as 
$t_{ij}=t_{0}a/r_{ij}$. $H_{\text{Au}}$ describes a band of ES. I will
discuss alternatively either bulk or Shockley surface states of the Au(111)
surface. $H_{V}$ is the hybridization between localized and ES, assuming
that the wave functions of the latter are plane waves \cite{revi}.

If one eliminates $H_{V}$ by means of a SWT {\em neglecting} $t_{ij}$ and
uses another canonical transformation to map the half filled $H_{\text{Cr}}$
into a Heisenberg model, the resulting effective Hamiltonian becomes
analogous to previously used ones \cite{kuda,laza}. As shown below, this
approach misses some coherence effects that reduce $T_{K}$ strongly 
for the equilateral trimer.

I solve exactly $H_{\text{Cr}}$ retaining only the ground state doublet in
the subspace of three particles $|3\sigma \rangle $, and then I perform a
generalized SWT to eliminate $H_{V}$. The validity of the approach is
discussed below. The resulting effective Hamiltonian has the form of a
one-impurity Kondo model

\begin{equation}
H_{K}=H_{\text{Au}}+\sum_{{\bf kk}^{\prime }\sigma \sigma ^{\prime }}J_{{\bf %
kk}^{\prime }}c_{{\bf k}\sigma }^{\dagger }\frac{{\boldmath \sigma }_{\sigma
\sigma ^{\prime }}}{2}c_{{\bf k}^{\prime }\sigma ^{\prime }}\cdot {\bf S,}
\label{hk}
\end{equation}
where ${\bf S}$ describes the spin operators of the doublet $|3\sigma
\rangle $ and ${\boldmath \sigma }_{\sigma \sigma ^{\prime }}$ is a vector
with the three Pauli matrices. The effective exchange can be written in the
form

\begin{eqnarray}
J_{{\bf kk}^{\prime }} &=&\bar{V}_{{\bf k}}V_{{\bf k}^{\prime }}\sum_{\alpha
}\{(\frac{1}{E_{2s\alpha }-E_{3}+\epsilon _{{\bf k}}} \label{jkk} \nonumber \\
&&+\frac{1}{E_{2s\alpha }-E_{3}+\epsilon _{{\bf k}^{\prime }}})\bar{A}%
_{2s\alpha }^{{\bf k}\uparrow }A_{2s\alpha }^{{\bf k}^{\prime }\uparrow } 
\nonumber \\
&&+(\frac{1}{E_{4s\alpha }-E_{3}-\epsilon _{{\bf k}}}+\frac{1}{E_{4s\alpha
}-E_{3}-\epsilon _{{\bf k}^{\prime }}})\bar{B}_{4s\alpha }^{{\bf k}%
\downarrow }B_{4s\alpha }^{{\bf k}^{\prime }\downarrow }  \nonumber \\
&&-(\frac{1/2}{E_{2t\alpha }-E_{3}+\epsilon _{{\bf k}}}+\frac{1/2}{%
E_{2t\alpha }-E_{3}+\epsilon _{{\bf k}^{\prime }}})\bar{A}_{2s\alpha }^{{\bf %
k}\downarrow }A_{2s\alpha }^{{\bf k}^{\prime }\downarrow }  \nonumber \\
&&-(\frac{1/2}{E_{4t\alpha }-E_{3}-\epsilon _{{\bf k}}}+\frac{1/2}{%
E_{4t\alpha }-E_{3}-\epsilon _{{\bf k}^{\prime }}})\bar{B}_{4s\alpha }^{{\bf %
k}\uparrow }B_{4s\alpha }^{{\bf k}^{\prime }\uparrow }\},  \nonumber \\
&&  
\end{eqnarray}
where

\begin{eqnarray}
A_{\nu }^{{\bf k}\sigma } &=&\langle 3\uparrow |\sum_{j}e^{i{\bf k\cdot R}%
_{j}}d_{j\sigma }^{\dagger }|\nu \rangle ,\text{ }  \nonumber \\
B_{\nu }^{{\bf k}\sigma } &=&\langle \nu |\sum_{j}e^{i{\bf k\cdot R}%
_{j}}d_{j\sigma }^{\dagger }|3\uparrow \rangle ,  \label{ab}
\end{eqnarray}
$|ns\alpha \rangle $ ($|nt\alpha \rangle $) denote the six singlet
eigenstates (three triplet eigenstates with maximum spin projection) of $H_{%
\text{Cr}}$ with $n=$2, 4 particles, $E_{\nu }$ is the energy of the state $%
|\nu \rangle $, and $E_{3}$ is the energy of the ground state doublet $%
|3\sigma \rangle $.

The wave vector dependence of the effective exchange interaction $J_{{\bf kk}%
^{\prime }}$ is rather involved. If only one virtual excited state dominates
in Eq. (\ref{jkk}), approximating $\epsilon _{{\bf k}}\simeq \epsilon _{F}$,
where $\epsilon _{F}$ is the Fermi energy (taken as zero), the effective
exchange has the functional form $J_{{\bf kk}^{\prime }}=\bar{f}({\bf k})f(%
{\bf k}^{\prime })C$, where $C$ is constant. In this case, it is easy to see
that if a Yosida trial wave function \cite{yos} is used to calculate the
ground state energy, the result is the same if $J_{{\bf kk}^{\prime }}$ is
replaced by the angular average of $J_{{\bf kk}}$. Guided by this result I
calculate this average evaluated at the Fermi energy in the general case
(calling it simply $J$) to discuss the dependence of $T_{K}$ on geometry.
It will become apparent that the qualitative features are not affected by
this approximation. Except for some detail discussed below, I assume
constant $V_{{\bf k}}$. Then the expression for $J$ is the same as Eq. (\ref
{jkk}) with the replacements

\begin{eqnarray}
\bar{V}_{{\bf k}}V_{{\bf k}^{\prime }}\bar{A}_{\nu }^{{\bf k}\sigma }A_{\nu
}^{{\bf k}^{\prime }\sigma } &\rightarrow &V^{2}\{\sum_{i}(\tilde{A}_{j\nu
}^{\sigma })^{2}+2\sum_{i<j}\beta _{ij}\tilde{A}_{i\nu }^{\sigma }\tilde{A}%
_{j\nu }^{\sigma }\},  \nonumber \\
\bar{V}_{{\bf k}}V_{{\bf k}^{\prime }}\bar{B}_{\nu }^{{\bf k}\sigma }B_{\nu
}^{{\bf k}^{\prime }\sigma } &\rightarrow &V^{2}\{\sum_{i}(\tilde{B}_{j\nu
}^{\sigma })^{2}+2\sum_{i<j}\beta _{ij}\tilde{B}_{i\nu }^{\sigma }\tilde{B}%
_{j\nu }^{\sigma }\},  \label{rep}
\end{eqnarray}
and $\epsilon _{{\bf k}}\rightarrow \epsilon _{F}$, with $\tilde{A}_{j\nu
}^{\sigma }=\langle 3\uparrow |d_{j\sigma }^{\dagger }|\nu \rangle $, $%
\tilde{B}_{j\nu }^{\sigma }=\langle \nu |d_{j\sigma }^{\dagger }|3\uparrow
\rangle $, and $\beta _{ij}$ is the angular average of 
$\cos \left( {\bf k}\cdot ({\bf R}_{i}-{\bf R}_{j})\right) $ with $\epsilon _{{\bf k}}$ on the
Fermi surface. For an isotropic two-dimensional (2D) band 
$\beta_{ij}=J_{0}(k_{F}r_{ij})$, where $J_{0}(x)$ is the zeroth order Bessel
function and $k_{F}a\simeq 0.6$, The corresponding 3D result is $\beta
_{ij}=\sin (k_{F}r_{ij})/(k_{F}r_{ij})$, with $k_{F}a\simeq 3.5$. However,
free electrons with $\epsilon _{{\bf k}}=\epsilon _{F}$ and small wave
vector parallel to the surface (implying small ${\bf k}\cdot ({\bf R}_{i}-%
{\bf R}_{j})$) have a larger wave vector perpendicular to the surface.
Naturally, one expects a larger extension of these states out of the surface
and therefore a larger hybridization with the adatoms \cite{revi,lin}. To
simulate this effect I take $k_{F}a$ as a parameter smaller than or equal to
$3.5$ for the calculations in which bulk states are assumed to dominate the
hybridization with the localized states.

\begin{figure}
\hskip1.0cm\psfig{file=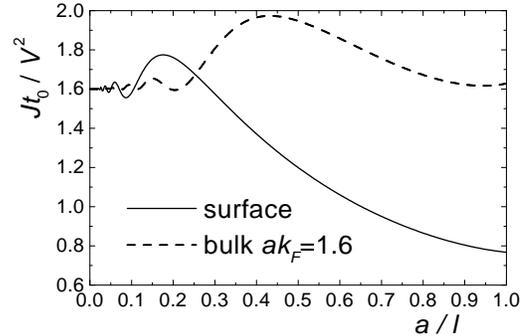,width=70mm,silent=} 
\caption{Average exchange interaction $J$ as a function of $1/l$ 
for an isoceles trimer with two sides of length $l$ and one side
of length $a$. Full (dashed) line correpond to surface 
(bulk with $ak_F=1.6$) ES. Other parameters are 
$E_d/t_0=-2.5$, $U/t_0=5$.}
\label{s1b1}
\end{figure}

In Fig. \ref{s1b1} I show the average exchange $J$ for an isosceles trimer
with $r_{23}=a$ and $r_{12}=r_{13}=l$, as $l$ is varied from $+\infty $ to $%
a $. I take $t_{23}=t_{0}$ as the unit of energy. $E_{d}$ and $U$ were
chosen to correspond to an electron-hole-symmetric situation for only one
adatom. Taking surface states as the ES, and disregarding small oscillations
coming from $J_{0}(x)$ at large argument, there is a moderate increase in $J$
as the Cr atom 1 approaches the dimer formed by Cr atoms 2 and 3, with
a maximum near $l=5a$. For smaller $l$, $J$ decreases monotonically until
the equilateral shape is reached. The situation is qualitatively similar if
the localized states are hybridized with bulk ES. The main difference is the
larger value of $k_{F}$. As $k_{F}$ is increased, the maximum is shifted to
smaller values of $l$ and larger values of $J$. For $k_{F}a=3.5$, the
average exchange always increases with decreasing $l$, except for the small
initial oscillations, while for $k_{F}a=0$, $J$ is a monotonically
decreasing function \cite{note2}.

To estimate the Kondo temperature for $k_{F}a=1.6$ (dashed line), I take a
density of states per spin and site $\rho =0.15/$eV, $t_{0}=V=0.5$ eV, a
band width $D=10^{5}$ K \cite{notep}, and use the expression \cite{wil}

\begin{equation}
T_{K}=D\sqrt{\rho J}\exp \left( \frac{-1}{\rho J}\right) .  \label{tk}
\end{equation}
This gives $T_{K}=8.3$ K for the single adatom ($l\rightarrow \infty $), $%
T_{K}=9.7$ K for the equilateral trimer, and a maximum $T_{K}=45$ K for $%
l=2.3a$. This behavior is roughly consistent with the experimental
observations \cite{jan}. While $T_{K}$ for the monomer and equilateral
triangle are larger than the experimental bound $T_{K}<7$, it is possible
that taking into account more precisely the wave vector dependence of $J_{%
{\bf kk}^{\prime }}$ improves the agreement. A scenario leading to a more
drastic change in $T_{K}$ is discussed below.

For $l\rightarrow \infty $, the average exchange coincides with the well
known result for a single adatom as expected \cite{sw}: 
$J=2V^{2}U/[(E_{d}+U)(-E_{d})]$. The variation of $J$ as $l$ decreases is the
result of mainly two competing effects: 
for $\beta _{12}=\beta _{13}\simeq 0$, 
where there are no coherence
effects between the atoms, the relevant matrix elements 
[Eqs. (\ref{rep})] increase,
while for $l=a$ and $\beta _{ij}=1$, they 
cancel by symmetry. Specifically, if $\beta _{12}=\beta _{13}$ is
not too close to 1, the largest contribution to $J$ comes from the virtual
state $|2sg\rangle $, the ground state of $H_{\text{Cr}}$ in the subspace of
two particles, which is a singlet even under the mirror plane of symmetry of
the isosceles triangle $m$ \cite{note3}. For $l\rightarrow \infty $, only
the $\tilde{A}_{j\nu }^{\sigma }$, $\tilde{B}_{j\nu }^{\sigma }$ with $j=1$
are different from zero.  As $l$ decreases, the sums in 
Eqs. (\ref{rep}) 
increase due to the increasing contribution of $\tilde{A}_{j\nu
}^{\sigma }$, $\tilde{B}_{j\nu }^{\sigma }$ for $j=2,3$. Also, the energy 
level of adatom 1 is pushed up towards the Fermi energy due to its
mixture with the bonding combination of adatoms 2 and 3 (the denominator $%
E_{2sg}-E_{3}+\epsilon _{F}$ decreases). Both effects tend to increase $J$.

However, there is a competing effect due to symmetry, as the compact
equilateral shape is approached when $k_{F}l$ becomes small. For $l=a$, the
Hamiltonian is invariant under the operations of the point group $C_{3v}$.
For $k_{F}\rightarrow 0$ ($\beta _{ij}\rightarrow 1$), the operator
entering Eqs. (\ref{ab}) becomes $\sum_{i}d_{i\sigma }^{\dagger }$, 
which is invariant under the point group operations. 
The state  $|2sg\rangle $ is also
invariant, while for $l\rightarrow a$, the ground state for three particles
becomes four-fold degenerate with two spin doublets belonging to the
two-dimensional representation $E$ of $C_{3v}$. The matrix element of 
$\sum_{i}d_{i\sigma }^{\dagger }$ between any of these four states and 
$|2sg\rangle $ vanishes by symmetry \cite{note}. Therefore, the contribution
of virtual states containing $|2sg\rangle $, which is the most important to $%
J$ when $\beta _{ij}\simeq 0$, vanishes when $\beta _{ij}\rightarrow 1$.
This explains the decrease of $J$ in Fig. \ref{s1b1} as $l\rightarrow a$ 
\cite{note2}.

The ground state for three particles $|3\sigma \rangle $ is even under $m$
for $l>a$ and odd for $l<a$. As a consequence of this jump in the symmetry
of the ground state, $J$ becomes much smaller for an isosceles triangle with
one side longer than the other two, including the linear trimer as a limit.
In particular, the contribution of $|2sg\rangle $ is negligible for $l<a$,
since all matrix elements between even and odd states are very small. This
agrees nicely with the fact that no FKA is observed for the linear trimer 
\cite{jan}. These symmetry arguments are independent of the approximation
made in the averaging over wave vectors.

The SWT is valid when the matrix
elements connecting the ground states with excited states ($V_{{\bf k}%
}A_{\nu }^{{\bf k}\sigma }$, $V_{{\bf k}}B_{\nu }^{{\bf k}\sigma }$) are
smaller than the corresponding denominators. In addition, since I retained
only the ground state of $H_{\text{Cr}}$ in the three-particle sector, the
effective matrix elements connecting different states in this subspace
should be smaller than the corresponding energy difference. For reasonable
values of $U$, the eight states with one particle per site are well
separated from those of higher energy. These eight states are degenerate if
the separation between any two adatoms is large. However, in the situations
of interest at least two adatoms interact, and then the spin quadruplet is
well separated from the two spin doublets and can safely be neglected.
Precisely at the equilateral shape, the two spin doublets are degenerate and
our approach is invalid. However a very small distortion towards an
isosceles shape is enough to recover the validity of the SWT. This is
because the distortion splits the spin doublets even and odd under $m$ and
the matrix elements between any even and any odd state are very small (they
vanish for $k_{F}a\rightarrow 0$). For the parameters used above to estimate 
$T_{K}$, a 1\% change in the length of one side suffices.

\begin{figure}
\hskip1.0cm\psfig{file=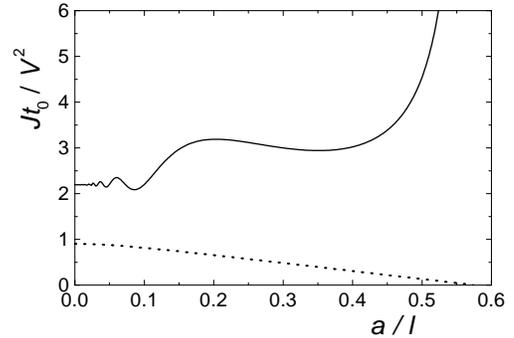,width=70mm,silent=} 
\caption{Same as Fig. 1 for surface ES and $E_d/t_0=-1.2$.
Dotted line is the difference $\epsilon_F+E_{2sg}-E_{3g}$.}
\label{s2}
\end{figure}

As mentioned before, the difference $E_{2sg}-E_{3}+\epsilon _{F}$ decreases
as $l$ decreases from $\infty $ to $a$, but is always positive and
significant for the parameters of Fig. \ref{s1b1}. However, for $E_{d}$
nearer to $\epsilon _{F}$ (for example $\epsilon _{F}-E_{d}<t_{0}$ if $U=0$)
there is a crossing point and the ground state of $H_{\text{Cr}}$ in the
equilateral trimer is the two-particle singlet $|2sg\rangle $. As a
consequence of the decrease in the corresponding denominator in $J$, there
is a dramatic increase in $T_{K}$ for decreasing $l$ near the crossing
point, until the system enters an intermediate valence regime (where the SWT
ceases to be valid) and for further decrease in $l$ the singlet dominates
the physics and there is no Kondo effect. The dependence of $J$ on $l$ in
a situation like this, taking surface states as the ES is illustrated in
Fig. \ref{s2}. The crossing point is at $a/l=0.576$. Taking $\rho
V^{2}/t_{0}=0.04$, the resulting Kondo temperature given by Eq. (\ref{tk})
is $T_{K}=0.33$ K for the single Cr adatom, has a relative maximum $T_{K}=14$
K at $a/l=0.2$, and after a minimum near 7K, reaches 44 K at $a/l=0.47$ and
174 K at $a/l=0.50$. However, at this point the quantitative validity of the
approach becomes questionable.

\begin{figure}
\hskip1.0cm\psfig{file=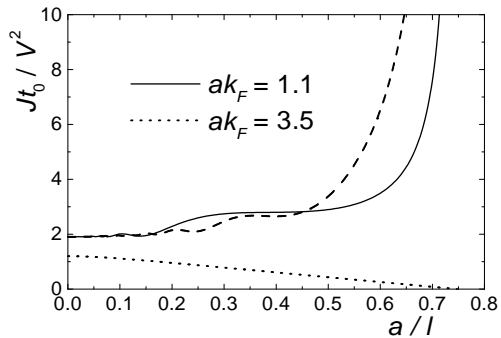,width=70mm,silent=} 
\caption{Same as Fig. 2 for bulk ES, $E_d/t_0=-1.5$
and two values of $k_F$.}
\label{b2b3}
\end{figure}

Fig. \ref{b2b3} shows a similar situation when bulk ES states dominate the
hybridization with the localized ones. The crossing point is at $a/l=0.75$.
The behavior of $J$ is more monotonic than in the previous case. Taking
again $\rho V^{2}/t_{0}=0.04$, one obtains $T_{K}=0.05$ K for $l\rightarrow
\infty $. For small $ak_{F}=1.1$, $T_{K}$ reaches 51 K at $a/l=0.47$ and 106
K at $a/l=0.55$. For $ak_{F}=3.5$, $T_{K}=50$ K at $a/l=0.52$ and 169 K at $%
a/l=0.55$, where the SWT is at the limit of its validity. Beyond the crossing
point ($a/l > 0.75$), no Kondo effect takes place.

In summary, I have constructed an effective Kondo model for a Cr trimer on a
Au(111) surface. For an isosceles trimer with one short side and two longer
sides of length $l$ and reasonable parameters, the model provides two
scenarios in which $T_{K}$ first increases with decreasing $l$ and then
decreases as the equilateral shape is approached. In the first one, this
decrease is due to symmetry effects and in the second one (which requires
smaller $\epsilon _{F}-E_{d}$), the ground state of the isolated equilateral
trimer is a singlet and no Kondo compensation of a localized spin takes
place. For an isosceles trimer with one side longer than the other two, or a
linear trimer, $T_{K}$ is very small due to the odd mirror symmetry of the
ground state. While these results are qualitatively consistent with
experiment \cite{jan}, a quantitative agreement with all the different
experimental observations seems to favor the second scenario, 
which is more robust under changes in the parameters,  and that bulk
states dominate the hybridization with the localized states of the Cr
adatoms. Since in the second scenario the electronic occupation of each Cr
atom in the compact equilateral trimer decreases by $\sim 1/3$, it is in
principle possible that X-ray experiments \cite{gam} which are sensitive to the Cr
valence can distinguish between the two scenarios. 

I thank A.M. Lobos, Karen Hallberg, P. Gambardella, J. Dorantes-D{\'a}vila 
and B. Normand for useful discussions. This work was sponsored by PICT 03-13829 of ANPCyT. I am
partially supported by CONICET.


\begin{references}
\bibitem{notep} The values of the parameters are guided by previous work
for Co impurities on Cu(111) [Ref. 8].
\bibitem{note}  This argument is easy to understand in the non-interacting
case $U=0$. The first two electrons occupy the invariant one-particle states 
$\sum_{i}d_{i\sigma }^{\dagger }|0\rangle $, with both spin projections, and
it is not possible to occupy a third state with the same symmetry due to
the Pauli principle.
\end{references}
\end{document}